\documentstyle[12pt]{article}
\input epsf
\title{\bf
Top-quark condensate at finite temperature and electroweak symmetry 
restoration
\thanks{This project was suppored partially by National Natural Science 
Foundation of China and by Grant No.LWTZ-1298 of the Chinese Academy of 
Sciences.}}
\author{
{\bf 
Bang-Rong Zhou} \\
\normalsize Department of Physics, Graduate School at Beijing,  
 University of Science and Technology \\
\normalsize of China, Academia Sinica, Beijing 100039, {\bf China} \\
\normalsize and  \\
\normalsize CCAST ( World Laboratory ) P.O.Box 8730, Beijing 100080,
{\bf China}\\
}
\date {}
\begin{document}
\hoffset = -1 truecm
\voffset = -2 truecm
\baselineskip = 12pt
\maketitle

\begin{abstract}
The gap equation at finite temperature in the top-quark condensate scheme of
electroweak symmetry breaking is proven to have the identical form in both the
imaginary and the real time formalism of thermal field theory.  By means of
the gap equation, combined with the basic relation to define the vacuum 
expectation value $v$ of the effective Higgs field, we analyse the dependence 
on temperature T and chemical potential $\mu$ of the dynamical top-quark mass 
as the order parameter characteristic of symmetry breaking,  and obtain 
the $\mu-T$ criticality curve for symmetry restoration.  We find out that 
the critical temperature $T_{c}=2v$ for $\mu=0$ and the critical chemical
potential $\mu_c=2\pi v/\sqrt{3}$ for $T=0$.  When $\mu=0$, the top-quark 
mass near $T_c$ has the leading ${(T_c^2-T^2)}^{1/2}$ behavior with
an extra factor dependent on temperature $T$ and the momentum cut-off
$\Lambda$. However, it is generally argued that the symmetry restoration at
$T\geq T_c$ is still a second-order phase transition. 
\end {abstract}
\section{Introduction}
Research on electroweak symmetry phase transition at finite 
temperature is attracting much attentions [1].  It has not only fundamental
field theory interest but also important practical implications to evolutiom 
of early universe and its cosmological results [2]. Many works have been made
in the scheme where symmetry breaking is induced by elementary Higgs field 
[3-7]. On the other hand, it is also quite interesting to discuss phase
transition of electroweak symmetry at high temperature in the scheme where
symmetry breaking is realized dynamically, i.e. by the condensates of 
fermions.  In view of that the top-quark condensate scheme [8,9] is a good 
candicate of such dynamical scheme at zero-temperature, in this paper we will
research its behavior at finite temperature. \\
\indent The top-quark condensate scheme is essentially a Nambu-Jona-Lasinio
(NJL) model [10].  Without loss of essentiality of such model we will made
our discussions only in the fermon bubble graph approximation and, to this 
order, completely neglect  color gauge 
interactions.  An obvious advantage of making this approximation is that the
whole discussions could be conducted analytically.  In this case, the only
order parameter responsible for electroweak symmetry breaking is the 
dynamical top-quark mass $m(T, \mu)$ as the function of temperature $T$ and 
chemical potential $\mu$, which is proportional to the thermal top-quark
condensate ${\langle\bar{t}t\rangle}_T$.  \\
\indent The paper is arranged as follows.  In Sect. 2 we will derive the gap 
equation at finite temperature and finite chemical potential respectively
from Feynman rules of the imaginary time formalism (ITF) and the real time 
formalism (RTF) of thermal field theory and show that 
both formalisms lead to the same result. In  Sect. 3  an equation for critical 
chemical potential $\mu_c$ and critical temperature $T_c$ is derived and  
the $\mu_c-T_c$ curve and some  numerical results of $\mu_c$ and $T_c$ are 
explicitly given.  In Sect. 4 
the temperature behavior of the dynamical quark mass $m(T, \mu)$ at 
$T\stackrel{<}{{}_{\sim}} T_c$ is discussed  and the phase transition's feature of 
the symmetry restoration is generally argued.  Finally, in Sect. 5 we briefly 
come to our conclusions. 
\section{Gap equation at finite temperature}  
In the top-quark condensate scheme at zero temperature, one assumes that the 
neutral scalar sector of the four-fermion interactions with the coupling 
constant $G$
$$
{\cal L}_{4F}^{N_S}=\frac{G}{4}{(\bar{t}t)}^2 
\eqno(2.1)$$
will lead to formation of the top-quark condensate $\langle\bar{t}t\rangle$ 
and generation of dynamical top-quark mass $m(0)$.  In the bubble graph 
approximation, the gap equation is [9]
$$m(0)=-\frac{G}{2}\langle\bar{t}(x)t(x)\rangle \eqno(2.2)$$ 
\noindent Equation (2.2) leads to 
$$m(0)=\frac{G}{2}d_t(R)\int \frac{d^4k}{{(2\pi)}^4}tr\left[\frac{i}
{\not\!{k}-m(0)+i\varepsilon}\right] \eqno(2.3)$$
\noindent and furthermore,
$$1=\frac{G\Lambda^2 d_t(R)}{8\pi^2}\left[
    1-\frac{m^2(0)}{\Lambda^2}\ln\left(\frac{\Lambda^2}{m^2(0)}+1\right)
    \right] \eqno(2.4)$$ 
\noindent where $\Lambda$ is the momentum cut-off in 4-dimension Euclidean
space and $d_t(R)$ is the dimension of the $SU_c(3)$ representation $R$ of 
the top-quark.\\
\indent The extension of Eq. (2.2) to finite 
temperature $T$ and finite chemical potential $\mu$ will be
$$m(T,\mu)=-\frac{G}{2}{\langle\bar{t}(x)t(x)\rangle}_T \eqno(2.5)$$
\noindent where ${\langle\bar{t}(x)t(x)\rangle}_T$ represents thermal expectation value.
However, since thermal field theory has two kinds of formalism: ITF and RTF 
[11], we must show that the extended gap equations (2.5) are equivalent in the 
two formalisms. In the following we will prove that they are in fact identical.
 \\
\indent 1) ITF: In this formalism the Feynman rules at zero-temperature 
should be modified according to that \\
fermion propagator:
$$\frac{i}{\not\!k-m(0)+i\varepsilon}\rightarrow
 \frac{\not\!k+m}{{(\omega_n+i\mu)}^2+{\it \bf k}^2+m^2} $$
$$ k^0=-i\omega_n+\mu, \ \ \ \  
\omega_n=\frac{(2n+1)\pi}{\beta}, \ \ \ \ n=0,\pm1, \pm2, \ldots 
\eqno(2.6)$$
\noindent with the denotations $\beta=1/T$ and $m\equiv m(T,\mu)$; \\
the momentum integral of a fermion loop:
$$\int \frac{d^4k}{{(2\pi)}^4} \rightarrow \int \frac{d^3k}{{(2\pi)}^3}
  T\sum_{n=-\infty}^{\infty} \eqno(2.7)$$ 
\noindent As a result, equation (2.5) can be changed into  
$$1=2Gd_t(R)\int \frac{d^3k}{{(2\pi)}^3}T\sum_{n=-\infty}^{\infty}
\frac{1}{{(\omega_n+i\mu)}^2+\omega_k^2}, \ \ \ \ 
 \omega_k^2\equiv \omega_k^2(T,\mu)={\it \bf k}^2+m^2
\eqno(2.8)$$
\noindent The sum in Eq. (2.8) can be done by means of the integral formula
$$T\sum_{n=-\infty}^{\infty}f[z=(2n+1)\pi /\beta+i\mu]
  =\frac{i}{4\pi}\oint_C dzf(z)\tan \left[\frac{1}{2}\beta(z-i\mu)\right] 
  \eqno(2.9)$$
\noindent where the contour $C$ contains the straight lines parallel to the 
real axis which extend from point $(i\mu-\infty)$ to $(i\mu+\infty)$ then goes 
back to the point $(i\mu-\infty)$ but round all the points 
$i\mu+(2n+1)\pi/\beta (n=0, \pm1,\pm2,\ldots)$.  By appropriately modifying 
the contour $C$, we will have that
\begin{eqnarray*}
\lefteqn{T\sum_{n=-\infty}^{\infty}f[z=(2n+1)\pi /\beta+i\mu]=} \\
 & & \frac{1}{2\pi}\int_{-\infty}^{\infty}dz f(z)-
 \frac{1}{2\pi}\int_{i(\mu-\varepsilon)-\infty}^{i(\mu-\varepsilon)+\infty}
  dz f(z){\left\{\exp[i\beta(z-i\mu)]+1\right\}}^{-1}+ \\
 & & \frac{1}{2\pi}\int_{i(\mu+\varepsilon)+\infty}^{i(\mu+\varepsilon)-\infty}
  dz f(z){\left\{\exp[-i\beta(z-i\mu)]+1\right\}}^{-1} 
\end{eqnarray*}$$\eqno(2.10)$$
\noindent where
$$f(z)=1/(z^2+\omega_k^2) \eqno(2.11)$$
Substituting Eq. (2.10) into Eq. (2.8), we obtain the gap equation at finite 
temperature
\begin{eqnarray*}
\lefteqn{1=2Gd_t(R)\left\{
    \int \frac{d^3k d\bar{k^0}}{{(2\pi)}^4}\frac{1}{{\bar{k^0}}^2+\omega_k^2}
                   \right. } \\  
  & & \left. - \int \frac{d^3k}{{(2\pi)}^3 2\omega_k}\left\{
   {\left\{\exp[\beta(\omega_k-\mu)]+1\right\}}^{-1}+
   {\left\{\exp[\beta(\omega_k+\mu)]+1\right\}}^{-1}\right\}
   \right\} 
\end{eqnarray*} $$\eqno(2.12)$$
\noindent Then by means of the Euclidean 4-dimension momentum cut-off 
$\Lambda$ Eq. (2.12) will be reduced to 
\begin{eqnarray*}
\lefteqn{1=\frac{G\Lambda^2d_t(R)}{8\pi^2}\left\{
    1-\frac{m^2}{\Lambda^2}\ln\left(\frac{\Lambda^2}{m^2}+1\right)\right.} \\
 & &\left.
   -\frac{4}{\Lambda^2}\int_{0}^{\infty}d|{\bf k}|\frac{|{\bf k}|^2}{\omega_k}
    \left\{{\left\{\exp[\beta(\omega_k-\mu)]+1\right\}}^{-1}+
           {\left\{\exp[\beta(\omega_k+\mu)]+1\right\}}^{-1}
    \right\} \right\}
\end{eqnarray*}  $$\eqno(2.13)$$
\noindent The 3-dimension momentum integral in the 
second term of right-handed side of Eq. (2.12) is convergent and this allows
us to have extended the upper limit of the integral to infinite. Such an
extension with the advantage of greatly simplifying the calculations seems
to have a little inconsistence with the fact that the four-fermion Lagrangian
(2.1) only express an effective interactions defined at the scales below the
compositeness scale $\Lambda$.  However, since the 3-dimension momentum 
integral from $\Lambda$ to infinity only contributes a very little part of the
total one, the extension can be viewed as a good approximation and will not
change the essential results.  When
$\mu=0$ and $\beta \rightarrow \infty (T\rightarrow 0)$, $m\equiv m(T,\mu)$
will become $m(0)$ and Eq. (2.13) will go back to the gap equation (2.4) in
zero-temperature case. \\
\indent 2) RTF:  In this formalism, all fields and interactions will be 
doubled.  The four-fermion interactions (2.1) should be replaced by
$$\bar{\cal L}^{N_S}_{4F}=\frac{G}{4}
\sum_{r=1,2}{(-1)}^{r+1}{(\bar{t}^{(r)}t^{(r)})}^2
  \eqno(2.14)$$
\noindent where $r=1$ means physical fields and $r=2$ ghost fields.  The 
physical fields and the ghost fields interact only through the propagator.
Thus the thermal propagator of the fermion will be a $2\times 2$ matrix:
$$i{\sf \bf S}(k,m)={\sf \bf M}\left(\matrix{
                      S(k,m) & 0           \cr
                      0      & S^{*}(k, m) \cr}
                \right){\sf \bf M} \eqno(2.15)$$
\noindent where $2\times 2$ matrix
$${\sf \bf M}=\left(\matrix{
         \cos\theta(k^0,\mu)  & -\exp[-\beta(k^0-\mu)/2]\sin\theta(k^0,\mu)\cr
     \exp[-\beta(k^0-\mu)/2]\sin\theta(k^0,\mu) & \cos\theta(k^0,\mu) \cr
        }\right) \eqno(2.16)$$
\noindent with
$$\sin\theta(k^0,\mu)=\frac{\theta(k^0)\exp[-\beta(k^0-\mu)/4]-
                            \theta(-k^0)\exp[\beta(k^0-\mu)/4]}
                           {{\left\{\exp[\beta(k^0-\mu)/2]+
                                  \exp[-\beta(k^0-\mu)/2]\right\}}^{1/2} }
                           \eqno(2.17)$$
\noindent and
$$S(k,m)=\frac{i}{\not\!k-m+i\varepsilon} \eqno(2.18)$$
\noindent Because the fermion mass $m$ is generated merely dynamically through
the thermal top-quark condensate ${\langle\bar{t}(x)t(x)\rangle}_T$, the thermal fermon 
propagator $iS^{ab}(k,m)$ will submit to the following equations
$$iS^{ab}(k,m)=iS^{ab}(k,0)+iS^{ac}(k,0)i(-1)^c\Sigma^c[iS^{cb}(k,m)]
    \ \ \ a,b,c=1,2                   \eqno(2.19)$$
\noindent where repeat of an index means its sum and $i(-1)^c\Sigma^c$ 
represents the contribution of the fermion loop connected to a $c$-type 
four-fermion vertex with
$$\Sigma^c=\frac{G}{2}d_t(R)\int\frac{d^4k}{{(2\pi)}^4}
           tr[iS^{cc}(k,m)]  \ \ \ \ ({\rm no \ sum \ of } \ c) \eqno(2.20)$$
\noindent By means of the matrix propagator (2.15), Eq. (2.19)  can be 
explicitly reduced to
\begin{eqnarray*}
\lefteqn{\left(\matrix{S(k,m) & 0       \cr
                 0      & S^*(k,m)\cr} \right)
  = \left(\matrix{S(k,0) & 0       \cr
                  0      & S^*(k,0)\cr}\right) } \\ \\ 
 &  & \mbox{} +\left(\matrix{S(k,0) & 0       \cr
                  0      & S^*(k,0)\cr}\right){\sf \bf M}
   i\left(\matrix{\Sigma^1 & 0        \cr
                   0       & \Sigma^2 \cr}\right)
   {\sf \bf M}\left(\matrix{S(k,m) & 0       \cr
                  0      & S^*(k,m)\cr}\right)
 \end{eqnarray*} $$\eqno(2.21)$$
\noindent The formula (2.20) gives that
$$\Sigma^1=mG2d_t(R)\int\frac{d^4k}{{(2\pi)}^4}\left[
           \frac{i}{k^2-m^2+i\varepsilon}-
           2\pi\delta(k^2-m^2)\sin^2\theta(k^0,\mu)\right]
           \eqno(2.22)$$
$$\Sigma^2={\Sigma^1}^* \eqno(2.23)$$
\noindent After counter clockwise and clockwise wick rotation of the integral
contour of $k^0$ respectively, it is easy to verify that $\Sigma^1$ and 
$\Sigma^2$ are both real and then from Eqs  (2.23) and (2.16) we may write the 
factor in Eq. (2.21)
$${\sf \bf M}i\left(\matrix{\Sigma^1  & 0          \cr
                   0        & \Sigma^2   \cr}\right){\sf \bf M}=
  -i\Sigma^1\left(\matrix{1    &  0    \cr
                          0    &  -1   \cr}\right) \eqno(2.24)$$
\noindent Substituting Eq. (2.24) into Eq. (2.21), we find out
$$S(k,m)=\frac{i}{\not\!{k}-\Sigma^1+i\varepsilon},  \ \ 
S^*(k,m)=\frac{-i}{\not\!{k}-\Sigma^1-i\varepsilon}  \eqno(2.25)$$
\noindent Hence, we may identify $\Sigma^1$ with the dynamical
fermion mass $m\equiv m(T, \mu)$ and then from Eq. (2.22) obtain 
$$1=G2d_t(R)\int \frac{d^4k}{{(2\pi)}^4}\left[
    \frac{i}{k^2-m^2+i\varepsilon}-2\pi\delta(k^2-m^2)\sin^2\theta(k^0,\mu)
    \right] \eqno(2.26)$$
\noindent which is the gap equation in RTF.  In fact, it is easy to 
see that Eq. (2.26) can be obtained from Eq. (2.3) simply through the replacement of 
the fermion propagator 
$$\frac{i}{\not\!k-m(0)+i\varepsilon}\rightarrow
  iS^{11}(k,m)=\frac{i}{\not\!k-m+i\varepsilon}-
               2\pi\delta(k^2-m^2)(\not\!k+m)\sin^2\theta(k^0, \mu)
  \eqno(2.27)$$
\noindent includung the replacement $m(0)\to m$. The replacement (2.27) is 
also valid in some other simple problems.  
However, only through the above derivation based on the matrix propagator
in RTF it can be believed that correctness of the replacement (2.27) has 
been proven rigorously.\\
\indent Equation (2.26) can be changed into the form of Eq. (2.13) after the 
Wick rotation, angular integration and introduction of the Euclidean 
4-dimension momentum cut-off $\Lambda$.  This indicates that the gap equation 
(2.5) are identical in ITF and RTF.  Furthermore,
we can eliminate the four-fermion coupling constant $G$ in Eq. (2.13)  by 
means of the zero-temperature gap equation (2.4) and transform Eq. (2.13) into 
\begin{eqnarray*}
\lefteqn{m^2(0)\ln\left(\frac{\Lambda^2}{m^2(0)}+1\right)=
  m^2\ln\left(\frac{\Lambda^2}{m^2}+1\right) }\\
 & & \mbox{}  +4\int_{0}^{\infty}d|{\bf k}|\frac{|{\bf k}|^2}{\omega_k}\left\{
  {\left\{\exp[\beta(\omega_k-\mu)]+1\right\}}^{-1}+
  {\left\{\exp[\beta(\omega_k+\mu)]+1\right\}}^{-1}\right\}
 \end{eqnarray*} $$\eqno(2.28)$$
\noindent which will be the starting equation in our following discussions. 
\section{The $\mu-T$ criticality equation and curve}  
For the zero-temperature top-quark condensate scheme of electroweak 
symmetry breaking in the fermion bubble graph approximation, we have the 
following basic relation [9]
$$\frac{4\sqrt{2}\pi^2}{3G_F}=\frac{8\pi^2}{3}v^2=
  m^2(0)\ln\left[\frac{\Lambda^2}{m^2(0)}+1\right] \eqno(3.1)$$
\noindent where $G_F$ is the Fermi constant and $v$ is the vacuum expectation 
value of the Higgs field in the standard electroweak model.  As a result of 
Eq. (3.1), equation (2.28) may be reduced to 
\begin{eqnarray*}
\frac{8\pi^2}{3}v^2&=&
  m^2\ln\left(\frac{\Lambda^2}{m^2}+1\right) \\
  & & \mbox{} +4\int_{0}^{\infty}d|{\bf k}|\frac{|{\bf k}|^2}{\omega_k}\left\{
  {\left\{\exp[\beta(\omega_k-\mu)]+1\right\}}^{-1}+
  {\left\{\exp[\beta(\omega_k+\mu)]+1\right\}}^{-1}\right\}   
 \end{eqnarray*} $$\eqno(3.2)$$
\noindent We notice that the four-fermion coupling constant $G$ and
the zero-temperature dynamical fermion mass $m(0)$ have disappeared from
Eq. (3.2) and they have been replaced by the vacuum expectation value $v$ of 
the Higgs field. \\ 
\indent It is seen from Eq. (3.2) that for a fixed chemical potential $\mu$, 
the second term in the right-handed side of the equation will be growing 
as the temperature $T$ is going up ($\beta$ is going down).  However, in 
order to keep the left-handed side of the equation to be a
constant, the first term in the right-handed side must be descending and 
eventually, one will arrive at a 
critical temperature $T_c$, where $m(T_c, \mu)=0$, i.e. 
${\langle\bar{t}t\rangle}_{T_c}=0$.  This means that electroweak symmetry will be restored
at the temperature $T\geq T_c$ when $\mu$ is fixed.  The similar situation
will happen for a fixed temperature $T$.  That is, we will have a critical
chemical potential $\mu_c$, and $m(T,\mu_c)=0$ for $\mu\geq \mu_c$ when $T$ 
is fixed.  The equation which determines $\mu_c$ and $T_c$ can be obtained by
setting $m=0$ in Eq. (3.2).  The result is 
$$\frac{2\pi^2}{3}v^2=
  T^2\int_{0}^{\infty}dx\left[
  \frac{x}{\exp(x-r)+1}+
  \frac{x}{\exp(x+r)+1}\right]   
  \eqno(3.3)$$
\noindent where the denotations $x=\beta |{\bf k}|$ and $r=\beta\mu$ have been
used. By means of the formula [12]
$$\int_{0}^{\infty}dx\frac{x^s}{\exp(x\mp r)+1}=
  \exp(\pm r)\Gamma(s+1)\Phi[-\exp(\pm r), s+1, 1 ] 
   \eqno(3.4)$$
\noindent with the definition of Lerch's transcendent function
$$\Phi(z,s,a)=\sum_{k=0}^{\infty}\frac{z^k}{{(k+a)}^s} \eqno(3.5)$$
\noindent  equation (3.3) becomes
$$\frac{2\pi^2}{3}v^2=T^2\left\{
  \exp(\mu/T)\Phi[-\exp(\mu/T), 2,1] +
  \exp(-\mu/T)\Phi[-\exp(-\mu/T), 2,1]
  \right\} \eqno(3.6)$$
In view of that when $\mu=0$, 
$\Phi[-\exp(-\mu/T),2,1] = \Phi[-1,2,1]=\pi^2/12$, Eq. (3.6) will lead to 
$$T_c=2v \ , \ \ \ \ {\rm when} \ \ \mu=0 \eqno(3.7)$$
\noindent On the other hand, when $T\rightarrow 0$, i.e. $r=\mu/T \to \infty$
if $\mu$ is fixed, we can rewrite the factor $T^2=\mu^2/r^2$ in Eq. (3.3) and
then, in terms of the limits 
$$\lim \limits_{r\to \infty}\frac{1}{r^2}\int_{0}^{\infty}dx
  \frac{x}{\exp(x\mp r)+1}=\left\{\matrix{\frac{1}{2} \cr
                                          0          \cr}
                            \right.  \eqno(3.8)$$ 
\noindent obtain from Eq. (3.3) that
$$\mu_c=\frac{2\pi}{\sqrt{3}}v, \ \ \ \ {\rm when} \ \ T=0 \eqno(3.9)$$
\noindent From Eq. (3.6), we may draw the general $\mu-T$ criticality 
curve which is displayed in \mbox{Figure 1.}  \\
\epsfxsize=12cm \epsfbox{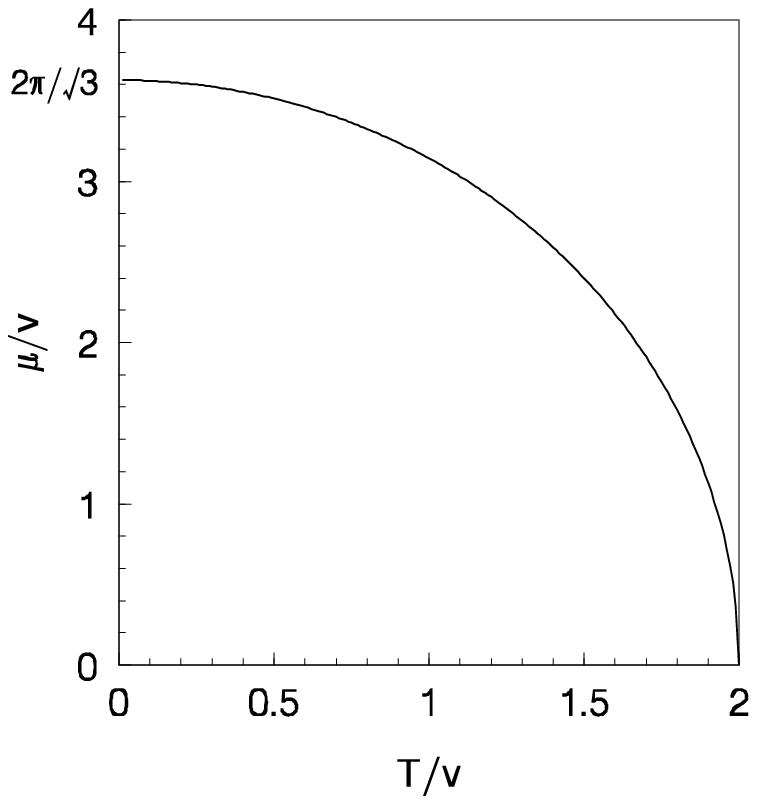}
\begin{center}
Fig.1 The critical chemical potential $\mu_c$-the critical temperature $T_c$
      curve \\ (scaled in the Higgs vacuum expectation value $v$).   
\end{center} 
\indent  In order to show numerical estimations of $\mu_c$ and $T_c$, we 
take $v=246 \ GeV$ and list the values of some special points in the 
$\mu_c-T_c$ curve in Table 1.  \\ 
\noindent Table 1.  Some special values of $\mu_c$ and $T_c$ when the Higgs
vacuum expectation value $v=246 \ GeV$ is taken.   \\ \\  
\begin{tabular}{lllllllllll}  \hline \\ 
$T_c(GeV)$       &0      &10     &50    &100   &300  
                 &400    &450    &485   &490   &492  \\    \hline \\ 
$\mu_c(GeV)$     &892.4  &892.2  &888   &874   &707   
                 &520    &361    &150   &80    &0      \\   \hline \\ 
\end{tabular} \\  
The expression (3.7) gives the highest critical temperature $T_c=2v=492 \ GeV$
which appears when $\mu=0$.  Such value of $T_c$ is higher than the 
temperature of electroweak symmetry restoration $T_c=\sqrt{2} v=347 \ GeV$
obtained in the theory with elementary Higgs scalar field if electroweak
gauge interactions are taken off ( and inclusion of these interactions will 
only make $T_c$ become lower.) [7].  This indicates a difference between 
dynamical and  non-dynamical scheme of electroweak symmetry breaking.
The formula (3.9) also shows that even if at $T=0$ the symmetry restoration could
still occur if the chemical potential $\mu \geq 2\pi v/\sqrt{3}=892 \ GeV$,
at least that is the case theoretically.  Generally, $T_c$ always 
monotonically decreases as $\mu$ increases.   
\section{Critical behavior of dynamical fermion mass}  
The critical behavior of the dynamical fermion mass $m\equiv m(T,\mu)$
at $T\stackrel{<}{{}_{\sim}} T_c$ can be obtained from Eq. (3.2) by high 
temperature expansion of the integrals in the equation.  For this purpose, we 
rewrite Eq. (3.2) as follows:
$$\frac{8\pi^2}{3}v^2=
  m^2\ln\left(\frac{\Lambda^2}{m^2}+1\right)+
  8T^2\left[I_3(y,-r)+I_3(y,r)\right] \eqno(4.1)$$
\noindent  where we have generally defined 
$$I_n(y,\mp r)=\frac{1}{\Gamma(n)}\int_0^{\infty}dx
\frac{x^{n-1}}{{(x^2+y^2)}^{1/2}}
\left\{\exp\left[{(x^2+y^2)}^{1/2}\mp r\right]+1\right\}^{-1} \eqno(4.2)$$
with the Gamma function $\Gamma(n)$ and the denotation $y=\beta m$.
By means of the recursion relations
$$\frac{dI_{n+1}(y, \mp r)}{dy}=-\frac{y}{n}I_{n-1}(y, \mp r)
  \eqno(4.3)$$
\noindent for $n>1$ and Eq.(3.4) we obtain 
\begin{eqnarray*}
\lefteqn{I_3(y,-r)+I_3(y,r)=
  -\frac{1}{2}\int_{0}^{y}dx x\left[I_1(x,-r)+I_1(x,r)\right]} \\
& & \mbox{} +\frac{1}{2}\left\{
                 \exp(r)\Phi[-\exp(r),2,1]+\exp(-r)\Phi[-\exp(-r),2,1]
              \right\} 
\end{eqnarray*}$$\eqno(4.4)$$
\noindent  On purpose to calculate $\left[I_1(x,-r)+I_1(x,r)\right]$, we make 
use of the formula
$$\frac{1}{e^{\theta}+1}=
  \frac{1}{2}-2\sum_{l=0}^{\infty}\frac{\theta}{{(2l+1)}^2\pi^2+\theta^2}
  \eqno(4.5)$$
\noindent and write that
\begin{eqnarray*}
\lefteqn{I_1(y,-r)+I_1(y,r) =
 \int_{0}^{\infty}\frac{dx}{{(x^2+y^2)}^{1/2}} } \\
& &\mbox{}-4\sum_{l=0}^{\infty}
\int_{0}^{\infty}\frac{dx}{x^2+{(2l+1)}^2\pi^2+y^2-r^2}
{\left\{1+\frac{4r^2{(2l+1)}^2\pi^2}{{[x^2+{(2l+1)}^2\pi^2+y^2-r^2]}^2}
\right\}}^{-1}  
\end{eqnarray*}  $$\eqno(4.6)$$
\noindent Assume that
$$\xi=\frac{r^2}{x^2+{(2l+1)}^2\pi^2+y^2}\leq \frac{r^2}{\pi^2} < 1
  \eqno(4.7)$$
\noindent and
$$\eta=\frac{y^2}{x^2+{(2l+1)}^2\pi^2}\leq \frac{y^2}{\pi^2} < 1
  \eqno(4.8)$$
\noindent then we can make the power series expansion of the integrand
in the second term of the right-handed side of Eq. (4.6) first in $\xi$ then
in $\eta$.  Retaining the terms up to the order $r^{\alpha}y^{\beta}$
with $\alpha+\beta\leq 6$ we will be able to write
$$I_1(y,-r)+I_1(y,r)=\lim \limits_{\varepsilon \to 0}\left(I_{1a}^{\varepsilon}+
I_{1b}^{\varepsilon}\right)+\sum_{i=2,4,6}f_iy^i+
                    \sum_{i=2,4,6}g_i(y^2)r^i \eqno(4.9)$$

\noindent   where 
$$I_{1a}^{\varepsilon}=\int_{0}^{\infty}dx\frac{x^{-\varepsilon}}
                       {{(x^2+y^2)}^{1/2}} \ \ {\rm and} \ \ 
I_{1b}^{\varepsilon}=-4\sum_{l=0}^{\infty}\int_{0}^{\infty}dx
         \frac{x^{-\varepsilon}}{x^2+{(2l+1)}^2\pi^2} \eqno(4.10)$$
\noindent are two divergent integrals which have been regularized by the 
factor $x^{-\varepsilon}(0<\varepsilon \ll 1)$  added in the integrands [5],
$f_i(i=2,4,6)$ are finite constants and $g_i(y^2)(i=2,4,6)$ are the
functions of $y^2$.  By means of the formula connected to the Euler beta 
function $B(p,q)$ [12]
$$\int_{0}^{\infty}dx\frac{x^{m-1}}{{(1+bx^a)}^{m+n}}=
  a^{-1}b^{-m/a}B\left(m/a,m+n-m/a\right) \ \ (a,b >0)
  \eqno(4.11)$$
\noindent it follows that
\begin{eqnarray*}
I_{1a}^{\varepsilon}&=&\frac{1}{2}y^{-\varepsilon}
   B\left(\frac{1-\varepsilon}{2},\frac{\varepsilon}{2}\right)=
   \frac{1}{2}y^{-\varepsilon}
   \Gamma\left(\frac{1-\varepsilon}{2}\right)
         \Gamma\left(\frac{\varepsilon}{2}\right)/
        \Gamma\left(\frac{1}{2}\right) \\
&=&\frac{1}{\varepsilon}-\ln\frac{y}{2}+O(\varepsilon) 
\end{eqnarray*}$$\eqno(4.12)$$
and in terms of the generalized Riemann zeta function $\zeta(s,a)$ and the 
Riemann zeta function $\zeta(s)$ [12] we obtain 
\begin{eqnarray*}
I_{1b}^{\varepsilon}&=&-\frac{1}{{(2\pi)}^{\varepsilon}}
  \zeta(1+\varepsilon,\frac{1}{2})=-\frac{1}{{(2\pi)}^{\varepsilon}}
  \left(2^{1+\varepsilon}-1\right)\zeta(1+\varepsilon) \\
  &=&-\frac{1}{\varepsilon}-\gamma+\ln\frac{\pi}{2}+O(\varepsilon) 
\end{eqnarray*}$$\eqno(4.13)$$
\noindent where $\gamma$ is the Euler constant.  As a result, the sum in 
Eq. (4.9)
$$\lim \limits_{\varepsilon \to 0}\left(I_{1a}^{\varepsilon}+
  I_{1b}^{\varepsilon}\right)=-\ln\frac{y}{\pi}-\gamma \eqno(4.14)$$
\noindent eventually become finite.  The other terms in Eq. (4.9) are all 
finite and can be expressed by the Riemann zeta functions $\zeta(s) (s=3,5,7)$.
Based on the resulting expression of $I_{1}(y,-r)+I_{1}(y,r)$  
and the result  from Eq. (4.4), the equation (4.1) finally becomes 
\begin{eqnarray*}
\lefteqn{ \frac{8\pi^2}{3}v^2 =
  m^2\ln\left(\frac{\Lambda^2}{m^2}+1\right)+
  4T^2\left\{ \exp(r)\Phi[-\exp(r),2,1]+\exp(-r)\Phi[-\exp(-r),2,1] 
  \right\} } \\
 & &\mbox{} +2m^2\left\{ \ln\frac{y}{\pi}-\frac{1}{2}+\gamma 
      -\frac{7}{4}\zeta(3)\left[{(y/2\pi)}^2+
       4{(r/2\pi)}^2\right] \right.   \\
 & &\mbox{}   +\frac{31}{8}\zeta(5)\left[ 
      {(y/2\pi)}^4+
      12{(y/2\pi)}^2{(r/2\pi)}^2+
      8{(r/2\pi)}^4 \right] \\
  & & \left.-\frac{127}{16}\zeta(7)\left[
      \frac{5}{4}{(y/2\pi)}^6+
      30{(y/2\pi)}^4 {(r/2\pi)}^2 +
      60{(y/2\pi)}^2 {(r/2\pi)}^4 +
      16{(r/2\pi)}^6 \right] \right\}   
\end{eqnarray*}$$\eqno(4.15) $$        
\noindent which is valid in the conditions (4.7) and (4.8).  We indicate
that , at the temperature $T\stackrel{<}{{}_{\sim}}T_c$, the condition (4.7) 
is easily
satisfied since in this case we will have $m\approx 0$ thus $y/\pi =m/{T\pi}
\ll 1$.  However, the condition (4.8) can not always be satisfied at a 
definite $T_c$.  This is because,  by observation of the $\mu-T$
criticality curve in Figure 1, for sufficient low $T_c$
we will have $\mu_c/T_c\pi \geq 1$. Therefore, the condition (4.8) means
that Eq. (4.15) is applicable only to the case of high $T_c$ where $\mu_c/
T_c\pi<1$ is satisfied, i.e. when $\mu \neq 0$, Eq. (4.15) can only be 
considered as the high temperature expansion of Eq. (4.1). \\
\indent The temperature dependence of $m\equiv m(T,\mu)$ at 
$T\stackrel{<}{{}_{\sim}} T_c$ can be derived from Eq. (4.15).   Since 
$m\approx 0$ when $T\stackrel{<}{{}_{\sim}} T_c$,
we will be able to omit the terms of the order ${(m/T)}^4 \ (y^4)$ and above 
in Eq. (4.15).  In addition, if $\mu/T\pi<1$ is assumed and 
$\Lambda^2/m^2\gg 1$ is considered, then Eq. (4.15) will be reduced to 
\begin{eqnarray*}
\frac{8\pi^2}{3}v^2 &=&
  4T^2\left\{ \exp(r)\Phi[-\exp(r),2,1]+\exp(-r)\Phi[-\exp(-r),2,1] 
      \right\}  \\
 & &\mbox{}+ 2m^2\left[
               \ln(\Lambda/T\pi)-\frac{1}{2}+\gamma  
               -7\zeta(3){(r/2\pi)}^2
               +31\zeta(5){(r/2\pi)}^4 
               -127\zeta(7){(r/2\pi)}^6   
         \right]      
\end{eqnarray*} $$\eqno(4.16)$$
\noindent Considering the equation (3.6) which determines the critical 
temperature $T_c$ (in which $T$ should be replaced by $T_c$) we may change 
Eq. (4.16) into 
\begin{eqnarray*}
\lefteqn{m^2=2\left[T_c^2\left\{ 
  \exp(r_c)\Phi[-\exp(r_c),2,1]+\exp(-r_c)\Phi[-\exp(-r_c),2,1] \right\}
     \right.}  \\ 
      & &  \left.-T^2\left\{ 
       \exp(r)\Phi[-\exp(r),2,1]+\exp(-r)\Phi[-\exp(-r),2,1] \right\}\right]/
     \\ 
      & &  \left[\ln(\Lambda/T\pi)-\frac{1}{2}+\gamma  
               -7\zeta(3){(r/2\pi)}^2
               +31\zeta(5){(r/2\pi)}^4 
               -127\zeta(7){(r/2\pi)}^6   
           \right]
\end{eqnarray*}$$\eqno(4.17)$$
\noindent where the denotation $r_c=\mu/T_c$ has been used.  The equation 
(4.17) gives the temperature dependence of $m$
at $T\stackrel{<}{{}_{\sim}} T_c$ when $\mu<T\pi$.  For the special case 
with $\mu=0$, it can be reduced to 
$$m=b(T){(T_c^2-T^2)}^{1/2}, \ \  b(T)=\pi /
\sqrt{3}{\left(\ln \frac{\Lambda}{T\pi}-\frac{1}{2}+\gamma\right)}^{1/2} 
\eqno(4.18)$$
\noindent By Eq. (2.5), the dynamical mass $m$ is proportional to the fermion
condensate ${\langle\bar{t}t\rangle}_T$ at temperature $T$ and  can be regarded as the
order parameter responsible for electroweak symmetry breaking.  Therefore,
the order parameter $m$ has the leading temperature behavior 
${(T_c^2-T^2)}^{1/2}$ at $T\stackrel{<}{{}_{\sim}} T_c$ , which is the same 
as the temperature behavior of the order parameter in the theory with  
elementary Higgs scalar field [7], but $m$ now contains an additional factor 
dependent on $\ln(\Lambda/T\pi)$.  For the case with non-zero chemical 
potential $\mu\neq 0$, the temperature behavior 
of $m$ at $T\stackrel{<}{{}_{\sim}}T_c$ should be determined by Eq. (4.17). \\ 
\indent Now let us return the case with $\mu=0$.
It may be generally shown that the temperature behavior ${(T_c^2-T^2)}^{1/2}$
of the order parameter m indicates that the symmetry restoration at 
$T\geq T_c$ will be a second-order phase transition in spite of the extra 
factor containing $\ln (\Lambda/T\pi)$ in $m$. This conclusion is the same as 
the one reached in the model with elementary Higgs fields [7].  In fact,
for a given model of fermion condensate, the momentum cut-off
or compositeness scale $\Lambda$ is a fixed finite constant.  In addition, 
since $T_c \neq 0$, the extra factor containing $\ln(\Lambda/T\pi)$ in $m$ 
will have no singularity at $T=T_c$.  Consequently, the feature of the phase 
transition will be determined completely by the temperature behavior 
${(T_c^2-T^2)}^{1/2}$ of the order parameter $m$.  Since $m$ is the unique 
order parameter for electroweak symmetry breaking in the scheme, we may 
expand the thermodynamical potential
$\Omega(T,m)\equiv \Omega(T,\mu=0, m)$ into power-series of $m$ near the 
critical temperature (phase transition point) as [13]
$$\Omega(T,m)=\Omega_0(T)+a(T-T_c)m^2+C(T)m^4+\cdots
\eqno(4.19)$$
\noindent where $a$ is a positive constant.  In Eq. (4.19), no term containing $m$
and $m^3$ appears.  This is out of the consideration of electroweak
symmetry.  Since in an effective Higgs field theory of the fermion condensate
scheme [9], the order parameter $m$ will be proportional to the vacuum
expectation value of the neutral composite Higgs field, the terms in the
effective Lagrangian which could lead to $m$ and $m^3$ term in Eq. (4.19)
have been removed by the $SU_L(2)\times U_Y(1)$ symmetry.   
The coefficient of $m^2$ term will ensure spontaneous
breaking at $T<T_c$ and symmetry restoration at $T>T_c$.  In addition, 
to make the phase transition point become a stable minimum of $\Omega(T,m)$
at $m=0$, it is necessary that the coefficient $C(T)>0$. \\
\indent From $\partial \Omega(T,m)/\partial m=0$ we obtain that when 
$T\leq T_c$ the order parameter $m$ can be expressed by
$$m^2=\frac{a}{2C(T)}(T_c-T) \eqno(4.20) $$
\noindent which will be identical to Eq. (4.18) if we identify
$$\frac{a}{2C(T_c)}=b^2(T_c)2T_c \eqno(4.21)$$
\noindent where the $T$s in $b(T)$, $C(T)$ and $(T_c+T)$ have been replaced 
approximately by $T_c$.  This implies that the thermodynamics of the system 
with order parameter (4.18) can be described by the thermodynamical potential
of the form (4.19).  Obviously, the entropy given by Eq. (4.19) (to $m^2$ 
order)
$$S=-{[\partial \Omega(T,m)/\partial T]}_V
 =\left\{\matrix{ 
         S_0-2aT_cb^2(T_c)(T_c-T) & {\rm if} \ \ T\leq T_c \cr
         S_0                      & {\rm if} \ \ T\geq T_c  \cr}
  \right.  
\eqno(4.22)$$
\noindent with 
$$S_0=-{[\partial \Omega_0(T)/\partial T]}_V \eqno(4.23)$$ 
and it will vary continuously when $T$ acrosses $T_c$.  But the special heat 
$$c_v=T{(\partial S/\partial T)}_V=
\left\{\matrix{
       c_{v0}+2aT_c^2b^2(T_c) & {\rm if} \ \ T < T_c \cr
       c_{v0}                 & {\rm if} \ \ T > T_c \cr}
\right.
\eqno(4.24)$$
\noindent  will experience a jump when $T$ acrosses $T_c$.  This verifies that
the phase transition is a second-order's one indeed.  We indicate that the 
extra factor containing $\ln(\Lambda/T\pi)$ in $b^2(T_c)$ does not change the 
second-order feature of the phase transition and at most, it simply affects 
the change rate of themodynamical quantities.       
\section{Conclusions}  
We have shown that a direct analysis of the dynamical fermion mass 
$m(T,\mu)$ is a quite effective approach for the  discussions of electroweak 
symmetry phase transition.  By means of the gap equation at finite temperature,
which is proven to be identical in both the 
imaginary and the real time formalism of thermal field theory,  incorporated 
with the basic relation to define the Higgs vacuum 
expectation value in the bubble graph approximation of the top-quark 
condensate scheme,  we have given a definite curve of the critical chemical 
potential $\mu_c$ and the critical temperature $T_c$ for electroweak symmetry
restoration and calculated explicit numerical values of $\mu_c$ and $T_c$.  
We have also proved by high temperature expansion of the thermal integrals 
that, for instance, when $\mu=0$  the order parameter $m(T,\mu)$ has the 
leading ${(T_c^2-T^2)}^{1/2}$ behavior but with an additional factor dependent
on $\ln (\Lambda/T\pi)$ at $T\stackrel{<}{{}_{\sim}} T_c$. However, despite of
the extra factor, the electroweak symmetry restoration at $T\geq T_c$ is still
a second-order phase transition.  This conclusion seems to show the 
essentiality of the scheme.  Certainly, whether it could be changed after the 
full dynamics of the gauge bosons and the composite Higgs boson are considered
may be researched further.  

\end{document}